\documentclass[12pt]{article}
\usepackage{amsmath,amssymb,amscd}
\usepackage{epsfig}
\begin{document}

\begin{center}
{\bf MARKET MILL DEPENDENCE PATTERN IN THE STOCK MARKET: MODELING OF PREDICTABILITY AND ASYMMETRY VIA
MULTI-COMPONENT CONDITIONAL DISTRIBUTION}
\end{center}

%\bigskip

\begin{center}
\bf{ \large Andrei Leonidov$^{(b,a,c)}$\footnote{Corresponding author. E-mail leonidov@lpi.ru}$^,$\footnote{Supported by the RFBR grant 06-06-80357},
Vladimir Trainin$^{(a)}$,
\\Alexander Zaitsev$^{(a)}$, Sergey Zaitsev$^{(a)}$}
\end{center}
\medskip

(a) {\it Letra Group, LLC, Boston, Massachusetts, USA}

(b) {\it Theoretical Physics Department, P.N.~Lebedev Physics Institute,\\
    Moscow, Russia}

(c) {\it Institute of Theoretical and Experimental Physics, Moscow, Russia}

%\bigskip

%\bigskip

\bigskip

\begin{center}
{\bf Abstract}
\end{center}

Recent studies have revealed a number of striking dependence patterns in high frequency stock price dynamics characterizing probabilistic
interrelation between two consequent price increments $x$ (push) and $y$ (response) as described by the bivariate probability distribution ${\cal
P}(x,y)$ \cite{LTZ05,LTZZ06a,LTZZ06b,LTZZ06c}. There are two properties, the market mill asymmetries of ${\cal P}(x,y)$ and predictability due to
nonzero $z$-shaped mean conditional response, that are of special importance. Main goal of the present paper is to put together a model reproducing
both the $z$-shaped mean conditional response and the market mill asymmetry of ${\cal P}(x,y)$ with respect to the axis $y=0$. We develop a
probabilistic model based on a multi-component ansatz for conditional distribution ${\cal P}(y|\,x)$ with push-dependent weights and means describing
both properties. A relationship between the market mill asymmetry and predictability is discussed. A possible connection
of the model to agent-based description of market dynamics is outlined.

\newpage

\section{Introduction}

This paper is based on recent results on high frequency conditional dynamics in the stock market \cite{LTZ05,LTZZ06a,LTZZ06b,LTZZ06c}. In
\cite{LTZZ06a,LTZZ06b,LTZZ06c} we have described several newly discovered dependence structures characterizing high frequency stock dynamics - the
market mill patterns corresponding to various asymmetries characterizing the bivariate probability distribution ${\cal P}(x,y)$ of two consecutive
price increments $x$ (push) and $y$ (response). We have also discussed a number of effects that are best described in terms of moments of the
conditional distribution ${\cal P}(y|\,x)$ of response $y$ at given push $x$. All the conditional moments studied (mean, standard deviation, skew,
normalized hypercumulant) reveal pronounced dependence on the push $x$. A nonzero z-shaped mean conditional response implying probabilistic
predictability of a price increment from knowing the previous one is of special interest. In this paper we will focus on building a probabilistic
model for both the nonlinear z-shaped mean conditional response and the corresponding market mill asymmetry pattern of ${\cal P}(x,y)$ with respect
to the axis $y=0$.

The issue of price predictability is perhaps the most important one in theoretical finance. In the ideally
efficient market it is impossible to predict future price increment using historical data, e.g values of the preceding
increments. If one describes price dynamics as a stochastic process, it should have zero conditional
mean $\langle y \rangle_x=0$, i.e. be a martingale, see e.g. \cite{M66,M71,S04}. Let us stress that an existence
of nonlinear dependence patterns related to higher moments of the conditional distribution does not contradict the
martingale property and thus the weak-form market efficiency. Such dependencies were intensively studied,
especially in the framework of conditional regressive dynamics \cite{S04,LB94}.

An appearance of the nonlinear mean conditional response shows that within a standard paradigm of regression models
formulated as a noisy mapping of the push $x$ into the response $y$, i.e. $y=f(x)+\varepsilon$, the mapping $f(x)=\langle
y \rangle_x$ is nonlinear. Thus we are dealing with a nonlinear dynamical system, see e.g. \cite{T90,D03}. In
financial applications a particular class of such models, the threshold autoregression models \cite{TL80} were
used in describing the properties of interest rates, see e.g. \cite{G05}.

At a fundamental level it is necessary to describe the nonzero nonlinear mean conditional response in terms of market inefficiency\footnote{Here we
refer to an idealized definition of market inefficiency in which the effects like bid-ask spread and trading costs limiting potential profit based on
specific predictability under discussion are discarded. } leading to probabilistic predictability. Some examples of such predictability were
discussed in \cite{M71}. Recently the issue of probabilistic predictability was discussed in the context of agent-based modeling of financial market
dynamics \cite{CG04}.

At a phenomenological level we need to build a probabilistic model explaining an origin of both the nonlinear
predictability of the mean conditional response and the corresponding market mill pattern within a simple and
intuitive probabilistic model. This is the main objective of the present paper.

When constructing a probabilistic model describing both the nonlinear conditional response and the market mill phenomenon we shall employ a
step-by-step approach. First, we provide an analytical description of the mean conditional response $\langle y \rangle_x$. Second, we consider a
simple dynamical system's version of noisy conditional dynamics characterized by the observed nonlinear mean conditional response dressed by an
additive noise. We show that this picture does not allow to reproduce the market mill asymmetry associated with conditional response. Finally, we
present a version of noisy conditional dynamics characterized by the push-dependent mixture of conditional distributions allowing to reproduce both
the nonlinear mean conditional response and the corresponding market mill asymmetry and suggest its interpretation in the framework of agent-based
modeling of financial markets.

The outline of the paper is as follows.

In Section 2 we start with describing in paragraph {\bf 2.1} basic quantitative characteristics of the asymmetry of the conditional response and an
algorithm allowing to reconstruct a full bivariate distribution ${\cal P}(x,y)$ from a given conditional distribution ${\cal P}(y|\,x)$. In paragraph
{\bf 2.2} we discuss a single-component conditional distribution corresponding to conventional noisy conditional dynamics that, by construction,
reproduces an observed nonlinear dependence of the mean conditional response. We show that this model gives rise to an asymmetry pattern very
different from the market mill one. In paragraph {\bf 2.3} a multicomponent conditional distribution of response at given push ${\cal P}(y|\,x)$ is
proposed. Its simple version with constant weights considered in paragraph {\bf 2.3.1} is shown to reproduce the market mill pattern but not the
nonlinear mean conditional response. Two versions with push-dependent weights described in paragraph  {\bf 2.3.2} are shown to produce both the
market-mill shaped asymmetry of the asymmetric component of ${\cal P}(x,y)$ and the nonlinearity of the conditional mean response $\langle y
\rangle_x$. In paragraph {\bf 2.4} we summarize the model-dependent relationships between the nonlinear predictability and the market mill pattern.
We proceed with comments on possible relation of the proposed probabilistic model to agent-based modeling of financial dynamics in paragraph {\bf
2.5} and conclude the section with a discussion of the limitations of the proposed probabilistic model in paragraph {\bf 2.6}.

In Section 3 we summarize the results of the present paper.

\section{Modeling the conditional response}

\subsection{General considerations}
At the basic level a probabilistic interrelation of the push $x$ and the response $y$ is quantified by the bivariate distribution ${\cal P}(x,y)$. As
already mentioned in the Introduction, in the preceding papers \cite{LTZZ06a,LTZZ06b,LTZZ06c} we have described a number of interesting features of
${\cal P}(x,y)$. However, from the modeling standpoint it turns out convenient to start with specifying a conditional distribution ${\cal P}(y|\,x)$.
Note that knowledge of  ${\cal P}(y|\,x)$ and the marginal distribution ${\cal P}(x)$ fully specifies ${\cal P}(x,y)\equiv {\cal P}(y|\,x) \, {\cal
P}(x)$. In the considered case of equal time intervals the marginal distributions ${\cal P}(x)$ and ${\cal P}(y)$ are the same. Thus for given
conditional distribution ${\cal P}(y|\,x)$ the marginal distribution ${\cal P}(x)$ should satisfy the following consistency equation
\begin{equation}\label{consrel}
{\cal P}(y) \, = \, \int dx \, {\cal P}(x) \, {\cal P}(y|\,x)
\end{equation}

The adopted procedure of constructing the bivariate distribution ${\cal P}(x,y)$ starting from a given ${\cal P}(y|\,x)$ goes in two steps:
\begin{itemize}
 {\item Constructing the conditional distribution ${\cal P}(y|\,x)$.}
 {\item Specifying  a reasonable marginal distribution ${\cal P}(x)$ satisfying the consistency condition (\ref{consrel}).}
\end{itemize}

In the present paper we shall focus on building a  model reproducing two phenomena described in \cite{LTZZ06a,LTZZ06b,LTZZ06c}:
\begin{itemize}
 {\item Market mill asymmetry of ${\cal P}(x,y)$ with respect to the axis $y=0$ \, .}
 {\item Predictability of response at given push due to the nonlinear $z$ - shaped mean conditional
 response $\langle y \rangle_x$ \, .}
\end{itemize}
These effects are illustrated in Figs.~1,2 correspondingly.

Both above-described features originate from the nontrivial asymmetric component ${\cal P}^a (x,y)$ of the distribution ${\cal P}(x,y)$:
\begin{equation}
{\cal P}^a \, = \, \frac{1}{2} \left( {\cal P} (x,y) - {\cal P}(x,-y)  \right )
\end{equation}
The market mill dependence pattern refers to the specific shape of the positive component ${\cal P}^{a (p)} (x,y)
\equiv {\cal P}^a (x,y) \cdot \Theta \left[ {\cal P}^a (x,y) \right]$ of ${\cal P}^a (x,y)$, where $\Theta$ is a
Heaviside step function. In turn, an emergence of the nonlinear $z$-shaped mean conditional response $\langle y
\rangle_x$ is quantified by the corresponding properties of the asymmetric component  ${\cal P}^a(y|\,x)$ of the
conditional distribution ${\cal P}(y|\,x)\equiv {\cal P}(x,y)/{\cal P}(x)$:
\begin{equation}\label{condmeanpush}
\langle y \rangle_x \, = \, \int dy \, y \,  {\cal P}^a (y|\,x)
\end{equation}

\subsection{Single-component conditional distribution}\label{scomp}

Let us first consider the following single-component conditional distribution:
\begin{equation}\label{condym2}
{\cal P} (y|\,x) \equiv {\cal P} (y-m(x)) \, ,
\end{equation}
where  $ m(x) \equiv \langle y \rangle_x $ is the mean conditional response having characteristic
$z$ - shaped form \cite{LTZZ06a}. To be concrete let us parametrize the push dependence of the mean conditional
response in the case
where $x$ and $y$ correspond to consecutive price increments in 3-minutes time intervals as follows:
\begin{equation}\label{parammeanconresp}
 m(x) \equiv \langle y \rangle_x \, = \, \left[ 0.14 - 0.24 \, |\,x|^{\,\, 0.15}   \right ] \cdot x
\end{equation}
In Fig.~3 we plot the analytical approximation (\ref{parammeanconresp}) together with corresponding market data from  Fig.~1.

To completely specify the conditional dynamics (\ref{condym2}) we have to choose a functional form of the distribution ${\cal P} (y|\,x)={\cal
P}(\varepsilon)$, where $\varepsilon \equiv y-m(x)$. Our choice is a Laplace distribution ${\cal P}^L (\varepsilon) = (0.5/\sigma) \exp
(-|\,\varepsilon|/\sigma)$ with $\sigma= \, \$ \, 0.052$\footnote{This choice of $\sigma$ corresponds to a mean absolute price increment at 3-minutes
scale $\langle |\,\delta p \,| \rangle= \, \$ \, 0.052$ for the ensemble of stocks and time period used in  \cite{LTZZ06a}. }. This selection is
motivated by the tent-like shape of the logarithm of the the central part of probability distribution of high-frequency price increments, see e.g.
\cite{LTZZ06b}. The conditional distribution is thus given by ${\cal P} (y|\,x)={\cal P}^L (y-m(x))$.

After having specified the conditional distribution ${\cal P} (y|\,x)$ and the marginal distribution ${\cal P}(x)$ we have to check that this choice
is consistent, so that these distributions satisfy Eq.~(\ref{consrel}). We have numerically checked that Eq.~(\ref{consrel}) is indeed satisfied by $
{\cal P} (y-m(x))$ and ${\cal P}^L (x)$ with above-described parameters.  Thus the full bivariate probability distribution ${\cal P} (x,y)$ is
completely specified. Its antisymmetric component is plotted in Fig.~4 (a). We conclude that reconstructing the conditional dynamics as described by
the single-component distribution (\ref{condym2}) with appropriate push - dependent mean (\ref{parammeanconresp}) leads to the two-dimensional
asymmetry pattern very different from the market mill one in Fig.~2. Thus here we have a model with proper nonlinear mean conditional response but
with wrong asymmetry pattern.

Let us note that the probabilistic model (\ref{condym2},\ref{parammeanconresp}) corresponds to a dynamical system
with nonlinear mapping $x \Longrightarrow m(x)$ dressed by additive noise $\varepsilon$ with distribution ${\cal
P} (\varepsilon)$ and zero mean $\langle \varepsilon \rangle = 0$:
\begin{equation}\label{condym1}
 y = m(x) + \varepsilon
\end{equation}
We see that a description in terms of a dynamical system (\ref{condym1}) explains (by construction) the
predictability related to the nonlinear $z$-shaped dependence of the conditional mean response on the push but leads
to the push-response asymmetry structure very different from the observed market mill shape.

\subsection{Multi-component conditional distribution}\label{mcomp}

Let us now construct a conditional distribution ${\cal P}(y|\,x)$ ensuring the $z$-shaped mean conditional
response in a different fashion. From the analysis of \cite{LTZZ06a,LTZZ06b,LTZZ06c} we know that the asymmetry in
question is relatively weak, so that a dominant component of ${\cal P}(y|\,x)$ should be symmetric with respect to
the reflection $y \to -y$ and, correspondingly, its dominant peak should be at $y=0$. In addition to the symmetric
component ${\cal P}^0 (y|\,x)$ having a constant weight $w^0$ the distribution should include two asymmetric
components  ${\cal P}^\pm (y|\,x)$ having push-dependent weights $w^\pm (x)$:
\begin{equation}\label{threemoddist}
{\cal P}(y\,|\,x) \, = \, w^+(x) \, {\cal P}^+(y\,|\,x)+ w^0 \, {\cal P}^0(y\,|\,x)+w^-(x) \, {\cal P}^-(y\,|\,x)
\end{equation}
where $w^+(x)+w^0+w^-(x)=1$. Below we shall take $w^0=0.85$. The distributions ${\cal P}^\pm (y|\,x)$ have peaks
at $y^\pm_*$ such that ${\rm sign} (y^\pm_*) = \pm {\rm sign} (x)$, so that the component ${\cal P}^+ (y|\,x)$
corresponds to a trend-following response, whereas the component ${\cal P}^- (y|\,x)$ corresponds to a contrarian
one. Below we shall use a simple parametrization of $m^\pm(x)$ ensuring these properties:
\begin{eqnarray}\label{condmeans}
 \langle y \rangle_x^+ \, \equiv \, m^+(x) & = &  (1+q) \, x \nonumber \\
 \langle y \rangle_x^0 \, \equiv \, m^0(x) & = &  0 \nonumber \\
 \langle y \rangle_x^- \, \equiv \, m^-(x) & = &  -(1-q) \, x \, ,
\end{eqnarray}
where $\langle y \rangle_x^+ \equiv \int dy \, y \, {\cal P}^+(y\,|\,x)$, and $q$ is a parameter responsible for generating the asymmetry in
question, for which we will take a value of $q=0.15$\footnote{The values of parameters contained in the model distribution ${\cal P}(y|\,x)$ in the
expressions for weights $w^\pm$ and parameters of the component distributions are fixed by fitting the resulting mean conditional response to the
market data corresponding to the time scale of 3 minutes.}. In Fig.~5 we plot the full conditional distribution ${\cal P} (y |\, x=0.1 )$ and its
three components using the weights $w^0=0.85$, $w^+=0.015$ and $w^-=0.135$. Let us note that a direct observation of multi-modal structure based on
market data can be very difficult. As could be seen from Fig.~5, a presence of two asymmetric contributions leads to a very slight distortion of the
full conditional distribution. At the same time their impact on asymmetry-sensitive quantities like the mean conditional response or the market mill
asymmetry can be very pronounced.

A simple calculation shows that the mean conditional response corresponding to Eq.~(\ref{threemoddist}) takes the
following form:
\begin{equation}\label{modcondresp}
 \langle y \rangle_x = \left[(w^+(x)-w^-(x)) + q \, (w^+(x)+w^-(x)) \right] \, x
\end{equation}

\subsubsection{Constant weights $w^\pm$}

It is instructive to start with the case of constant weights $w^\pm$. From Eq.~(\ref{modcondresp}) we see that in
this case one gets a purely linear correlation between the conditional mean response and the push. To analyze the
two-dimensional asymmetry pattern one has to specify the functional form of the distributions ${\cal P}^{\pm,\,0}
(y|\,x)$. Let us assume that all three are Laplace distributions
\begin{equation}\label{dl1}
{\cal P}^{\pm,\,0}(y|\,m^{\pm,\,0}(x),\sigma) \, = \, \frac{1}{2 \sigma} \, {\rm exp} \left\{ - |y-m^{\pm,\,0}(x)|/\sigma \right\}
\end{equation}
with common width parameter $\sigma$. We choose $w^\pm=0.075$, $q=0.15$ and $\sigma=\, \$ \,0.052$. Using the
consistency condition Eq.~(\ref{consrel}) we have checked that the distribution (\ref{threemoddist}) with constant
weights $w^\pm$ indeed solves (\ref{consrel}) so that we can reconstruct the full two-dimensional distribution
${\cal P}(x,y)$ and its asymmetric component ${\cal P}^a$. The resulting asymmetry pattern is shown in Fig.~4 (b).
We see that it has a characteristic market mill shape. We conclude that a model based on the distribution
(\ref{threemoddist}) with constant weights $w^\pm$ does reproduce the market mill asymmetry pattern but does not
reproduce the nonlinear  $z$-shaped mean conditional response.

\subsubsection{Push-dependent weights $w^\pm (x)$}

The experimentally observed dependence does have a pronounced nonlinear z-shaped form, see Fig.~1 and
\cite{LTZZ06a}, so we have to correct our model in order to reproduce it.  In fact, to generate such nonlinear
dependence it is sufficient to consider a case of push-dependent weights $w^{\pm}(x)$. To reproduce the
$z$-shaped pattern of Fig.~1 we have to ensure a bias towards trend-following behavior, (i.e positive slope of
$\langle y \rangle_x$ vs the push) at small nonzero $x$ and a bias towards contrarian behavior (negative slope of
$\langle y \rangle_x$ vs the push) at large $x$. The components of the conditional distribution
Eq.~(\ref{threemoddist}) responsible for trend-following and contrarian behavior are obviously ${\cal P}^+(y|\,x)$
and ${\cal P}^-(y|\,x)$ correspondingly, so we have to choose some appropriate parametrization of the weights
$w^\pm$. A simple illustration leading to the conditional mean response shape qualitatively similar to that in
Fig.~1 is provided by
\begin{equation}\label{simpleweights}
w^\pm (x) \, = \, \frac{1-w^0}{2} (1 \mp |\,x|) \,\,\, \Longrightarrow \,\,\, \langle y \rangle_x \, = \, (1-w^0)
\, x \, (q-|\,x|)
\end{equation}
From Eq.~(\ref{simpleweights}) we see that the role of the parameter $q$ is in fixing the scale at which the
nonlinear mean conditional response $\langle y \rangle_x$ changes its sign.

To reproduce the shape of conditional mean response close to the observed one, see Fig.~1, we shall use a somewhat more complex parametrization of
the weights $w^\pm (x)$:
\begin{eqnarray}\label{weights}
 w^-(x) & = & {\rm min} \left(  w^a \,+ \left(1-w^0-w^a \right)  \left( \frac{|\,x|}{0.3} \right)^p  , \,\,\,1-w^0 \right) \nonumber \\
 w^+(x) & = & 1 - w^0 - w^-(x)
\end{eqnarray}
Choosing $w^0=0.85$, $w^a=0.05$, $q=0.25$, $p=0.5$  and $\sigma = \, \$ \, 0.052$ gives an alternative parametrization of conditional response, cf.
Eq.~(\ref{parammeanconresp}). The resulting asymmetry pattern is plotted in Fig.~4 (c) and the push dependence of the weights $w^\pm (x)$ is
illustrated in Fig.~6. Let us note that such fine details of the structure of the asymmetry of the empirical distribution ${\cal P} (x,y)$ shown in
Fig.~2 as the shape of equiprobability lines and the varying form of mill blades are, as seen in Fig.~4 (á), reproduced by our model . Let us stress
that a bias towards trend-following behavior at small pushes and towards contrarian behavior at large ones results from combination of $x$ -
dependent weights and means.

Let us also emphasize that although it is formally possible to rewrite the conditional dynamics described by the three-modal distribution with
push-dependent weights Eq.~(\ref{threemoddist}) in the form of Eq.~(\ref{condym2},\ref{condym1}) with some very intricate noise distribution ${\cal
P}(\varepsilon)$, this procedure looks extremely unnatural. In this sense the three-component conditional distribution
Eq.~(\ref{threemoddist}) presents a really different view on conditional
dynamics than the conventional Eq.~(\ref{condym1}).

Thus a model based on the conditional distribution (\ref{threemoddist}) with specially chosen push-dependent weights $w^\pm(x)$ allows to reproduce
both the nonlinear mean conditional response and the market mill asymmetry pattern.

The above-discussed procedure of constructing the conditional distribution ${\cal P}(y|\,x)$ is by no means unique. For example, ways of introducing
the response asymmetry can be different. In particular, in the above-discussed parametrization Eqs.~(\ref{condmeans},\ref{weights}) the asymmetry in
question was ascribed to an asymmetric structure of the means of its components ${\cal P}(y|\,x)^\pm$.

Let us now introduce another model of the same class which is also based on the three-component conditional distribution ${\cal P}(y|\,x)$. In this
model the weights $w^\pm$ are the same as in Eq.~(\ref{weights}), but an asymmetric dependence of the characteristics of the component distributions
is "shifted" from the means $m^\pm(x)$ to the widths $\sigma^\pm(x)$:
\begin{eqnarray}\label{parvar}
 m^+(x) & = & \,\,x, \,\,\,\,\,\,\,\,\,\,\, \sigma^+(x)=0.005+0.065 \, \left( \frac{\sqrt{|\,x|}}{0.3} \right)^{0.5} \nonumber \\
 m^0(x) & = & \,\,0, \,\,\,\,\,\,\,\,\,\,\,\, \sigma^0 = 0.03 \nonumber \\
 m^-(x) & = & -x, \,\,\,\,\,\,\,\,\,\, \sigma^-(x)=0.01+0.025 \, \left( \frac{\sqrt{|\,x|}}{0.3} \right)^{0.5}
\end{eqnarray}
The resulting asymmetry pattern is shown in Fig.~4 (d).  All the results obtained using the model of Eqs.~(\ref{weights},\ref{parvar}) are similar to
those obtained with Eqs.~(\ref{condmeans},\ref{weights}).

The models of Eqs.~(\ref{condmeans},\ref{weights}) and (\ref{weights},\ref{parvar}) are just two examples from a
long list of models that describe the market mill asymmetry with respect to the axis $y=0$ and the nonlinear mean
conditional response. Our choice was motivated by their transparent logical structure.

\subsection{Relationship between nonlinear predictability and market mill asymmetry}

At this point it is appropriate to summarize the relationship between the phenomena of nonlinear
predictability due to nontrivial push-dependent mean conditional response and the market mill asymmetry pattern
characterizing the asymmetry of the bviariate distribution ${\cal P}(x,y)$ with respect to the axis $y=0$. Our
considerations in the paragraphs \ref{scomp} and \ref{mcomp} have shown that this relationship is model -
dependent. In particular:
\begin{itemize}
 \item{The single-component model (\ref{condym2},\ref{parammeanconresp}) of conditional dynamics describes
 the nonlinear predictability (i.e. the $z$-shaped mean conditional response) but gives a wrong asymmetry pattern of
 ${\cal P}(x,y)$. Thus the nonlinear mean conditional response does not constitute a sufficient condition for the
 existence of the market mill asymmetry.}
 \item{The multi-component model  (\ref{threemoddist},\ref{condmeans})  with constant weights $w^\pm$ describes
 the market mill asymmetry pattern but not the nonlinear dependence of the mean conditional response on push.
 Thus the market mill asymmetry does not constitute a sufficient condition for the existence of the nonlinear
 mean conditional response.}
\end{itemize}

\subsection{Market mill from the agent-based perspective}

Let us now discuss possible origins of the market mill asymmetry in the framework of agent-based description
of financial market dynamics.

A direct link between the agent's strategies and price evolution is provided by the relation between the sign of
market orders and the resulting change in price, see e.g. \cite{FJ02}. In the discrete time formulation this
corresponds to a dependence of price increment $\delta p_t=p_{\, t+1}-p_t$ on cumulative sum of signed orders
$\Omega_t=V^+_t-V^-_t$ placed at time $t$, where $V^\pm$ is a volume of buy (+) and sell (-) orders. With the
simplest assumption of linear impact
\begin{equation}\label{linim}
\delta p_{\,t} \, = \, \frac{1}{\lambda} \, \Omega_t \,.
\end{equation}
Assuming constant proportionality coefficient $\lambda$, the probability distribution of price increments is a
rescaled version of the probability distribution of the signed volume:
\begin{equation}
{\cal P} (\Omega_t) \,\,\,\,\ \longrightarrow \,\,\,\,\, {\cal P} (\delta p_{\,t})
\end{equation}

The distribution ${\cal P} (\Omega_t)$ provides a probabilistic description of agent's strategies realized through
buying (selling) a certain number of stocks or just doing nothing at time $t$. Let us consider a simple case when
such trading decisions depend on the preceding price increment $\delta p_{\, t-1}$, so that
\begin{equation}
{\cal P}\left( \delta p_{\, t} \right) = \frac{1}{\lambda} \, {\cal P} \left( \Omega_t | \, \delta p_{\, t-1}
\right).
\end{equation}
Within this framework it is natural to classify agents into three groups, trend-following, contrarian and noise, characterized by probability
distributions ${\cal P}^+ \left( \Omega_t | \, \delta p_{\, t-1} \right)$, ${\cal P}^- \left( \Omega_t | \, \delta p_{\, t-1} \right)$ and ${\cal
P}^0 \left( \Omega_t | \, \delta p_{\, t-1} \right)$ correspondingly. The distributions ${\cal P}^{\pm,0} \left( \Omega_t | \, \delta p_{\, t-1}
\right)$ are biased in such a way that
 $$
    {\rm sign} \left( \, \langle {\cal P}^\pm \left( \Omega_t | \, \delta p_{\, t-1} \right) \rangle \, \right) =
    \pm {\rm sign} \left( \delta p_{\, t-1} \right)
 $$
and $\langle \, {\cal P}^\pm \left( \Omega_t | \, \delta p_{\, t-1} \right) \, \rangle =0$. The trend-followers are betting that the sign of the next
price increment is on average the same as that of the previous one, the contrarians bet on sign reversal and noise traders make random decisions.
Generically the yields $w^{\pm,\, 0}$ of the trend-following, contrarian and noise strategies depend both on the sign and magnitude of $\delta p_{\,
t-1}$. Thus
\begin{equation}\label{threemodstrat}
 {\cal P} \left( \Omega_t | \, \delta p_{\, t-1} \right) = \sum_{i=+,-,\, 0}  \, w^i (\delta p_{\, t-1}) \cdot
 {\cal P}^i \left( \Omega_t | \, \delta p_{\,t-1} \right)
\end{equation}
which is, due to (\ref{linim}), precisely the conditional distribution (\ref{threemoddist}). Therefore the three components of the conditional
distribution ${\cal P} \left( \Omega_t | \, \delta p_{\, t-1} \right)$ in Eq.~(\ref{threemodstrat}) (trend-following, contrarian and noise)
correspond to the three components of the conditional distribution ${\cal P}(y|\,x)$ in Eq.~(\ref{threemoddist}).

\subsection{Limitations of the model}

In the present paper we have focused on building a probabilistic model describing the market mill asymmetry pattern corresponding to one particular
asymmetry of ${\cal P}(x,y)$, that of reflection $y \to -y$. As shown in \cite{LTZZ06a}, the full empirical bivariate distribution is in fact
characterized by several asymmetry patterns having the market mill shape, e.g. that corresponding to the reflection with respect to the axis $y=x$.
We have checked that with the three-component conditional distribution Eq.~(\ref{threemoddist}) one can not reproduce the market mill pattern
corresponding to this last asymmetry. In fact, our model is tailored to describe only one particular asymmetry, that of conditional response.
Constructing a probabilistic description of the full asymmetry structure of ${\cal P}(x,y)$ remains a task for the future.

Let us also note that even with the asymmetry pattern under consideration the model conditional distribution (\ref{threemoddist}) does not allow to
reproduce all details of the empirically observed pattern.

\section{Conclusions}

Let us summarize the main results obtained in the paper:
\begin{itemize}
 \item{A probabilistic model based on the multi-component model for conditional distribution ${\cal P}(y|\,x)$ reproducing
 the nonlinear $z$-shaped mean conditional response and the market mill conditional response asymmetry pattern was constructed.}
 \item{We demonstrated that a single-component model corresponding to conventional noisy conditional dynamics with
 built-in $z$-shaped mean conditional response does not allow to reproduce the market mill conditional response
 asymmetry pattern. Thus an existence of the z-shaped mean conditional response does not imply the
 market mill asymmetry pattern.}
 \item{Consideration of the case of push-independent weights in the multi-component model conditional distribution
 ${\cal P}(y|\,x)$ showed that the market mill asymmetry pattern can coexist with the usual linear dependence of
 the mean conditional response on push. Thus an existence of the market mill asymmetry pattern does not imply
 the z-shaped mean conditional response.}
 \item{A possible link of the discussed probabilistic model with agent-based description of market dynamics was outlined.}
\end{itemize}

\begin{figure}[h]
 \begin{center}
 \leavevmode
 \epsfxsize=14cm
 \epsfbox{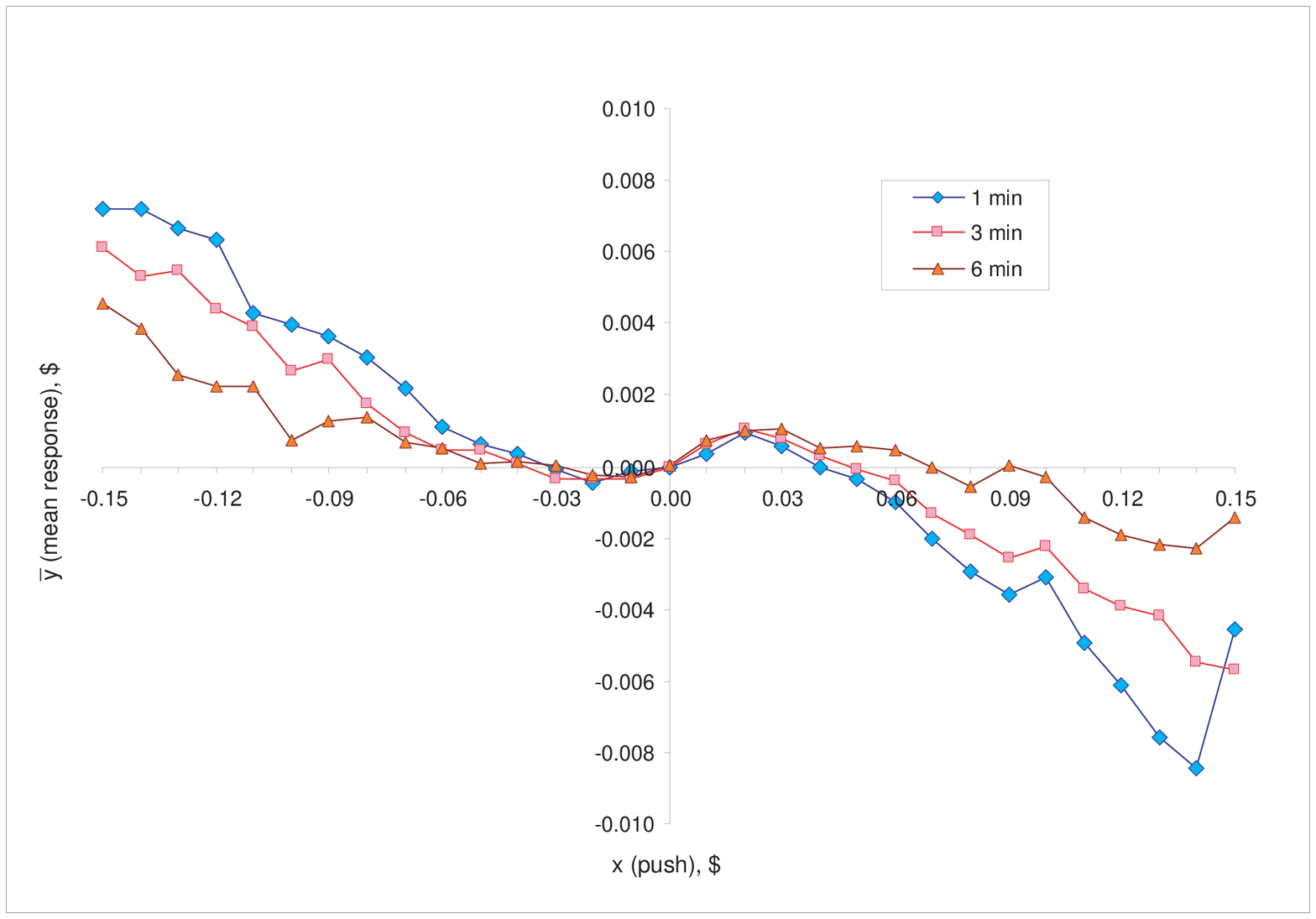}
 \end{center}
  \caption{Mean conditional response versus push (from \cite{LTZZ06a}).}
\end{figure}
\begin{figure}[h]
 \begin{center}
 \epsfig{file=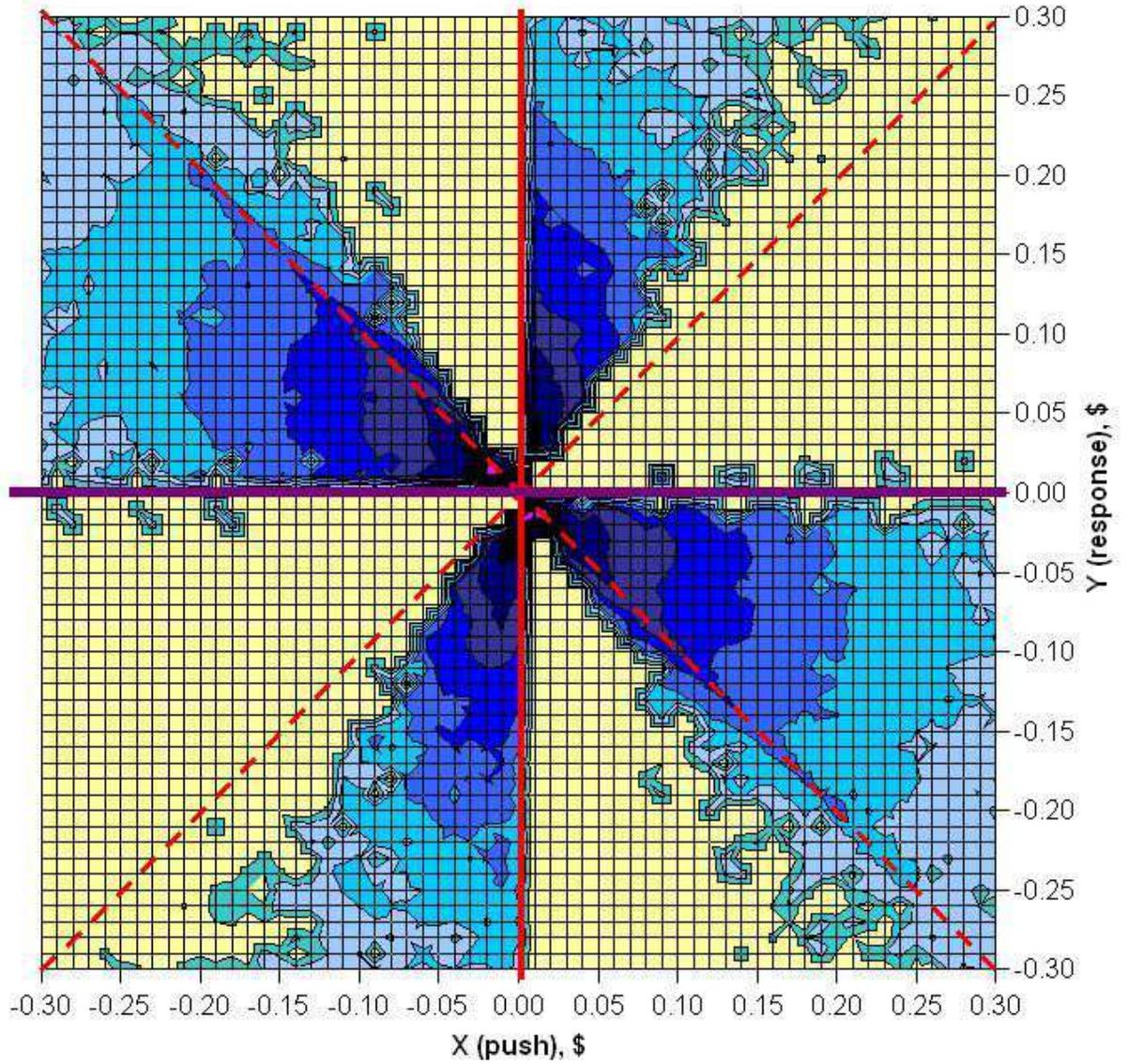,width=16cm}
 \end{center}
 \caption{Two-dimensional projection of the asymmetry of the bivariate distribution $\log_2 \left( {\cal P}(x,y) \right)$
 (from \cite{LTZZ06a}). Borders between different colors show equiprobability lines.}
\end{figure}
\begin{figure}[h]
 \begin{center}
 \leavevmode
 \epsfysize=10cm
 \epsfbox{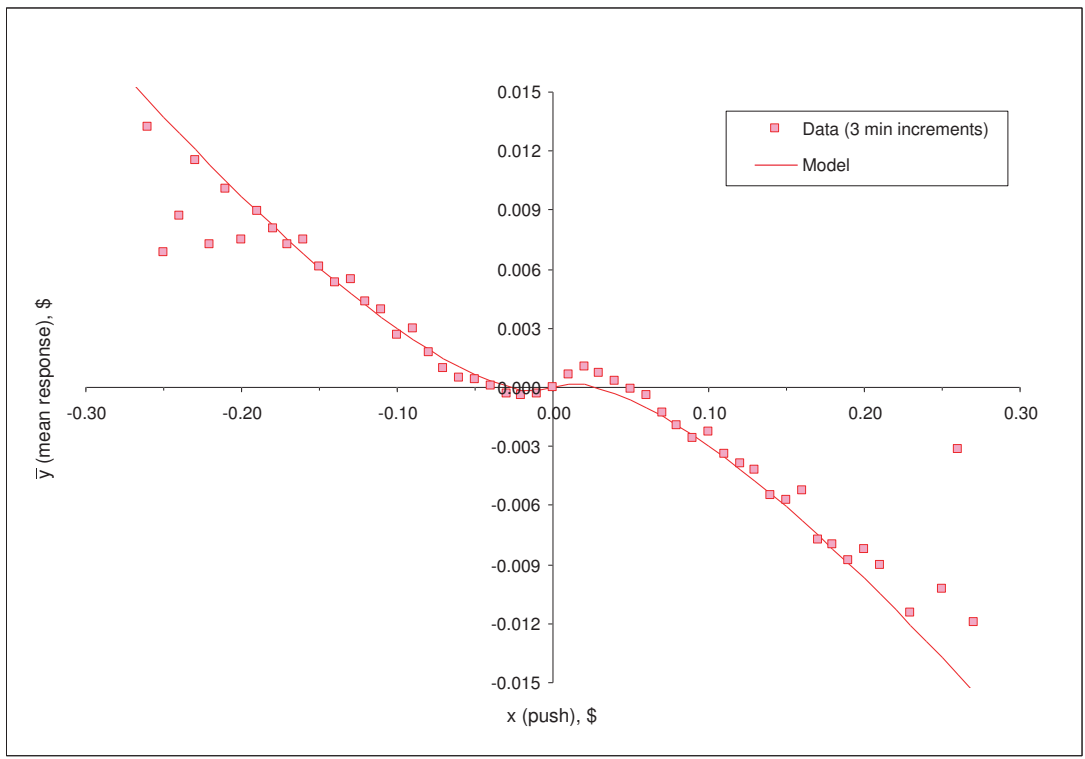}
 \end{center}
\caption{Mean conditional response: model versus data.}
\end{figure}
\begin{figure}[h]
 \begin{center}
 \leavevmode
 \epsfysize=18cm
 \epsfbox{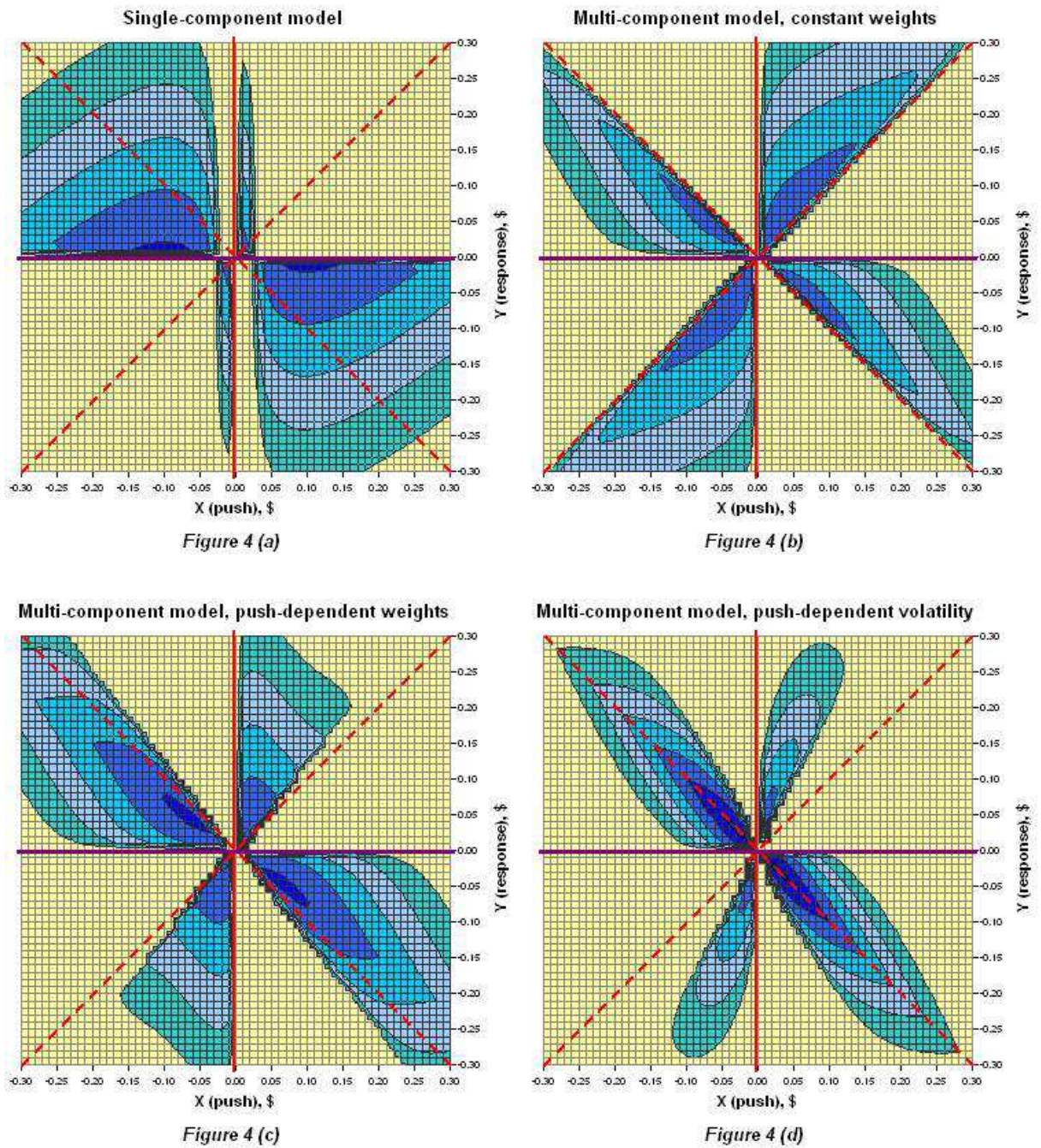}
 \end{center}
 \caption{Two-dimensional projection of the asymmetry of the bivariate distribution {\cal P}(x,y).
 (a) Single-component model. (b) Multi-component model, constant weights. (c) Multi-component model, push-dependent weights.
 (d) Multi-component model, push-dependent volatility. Borders between different colors show equiprobability lines.}
\end{figure}
\begin{figure}[h]
 \begin{center}
 \leavevmode
 \epsfysize=15cm
 \epsfbox{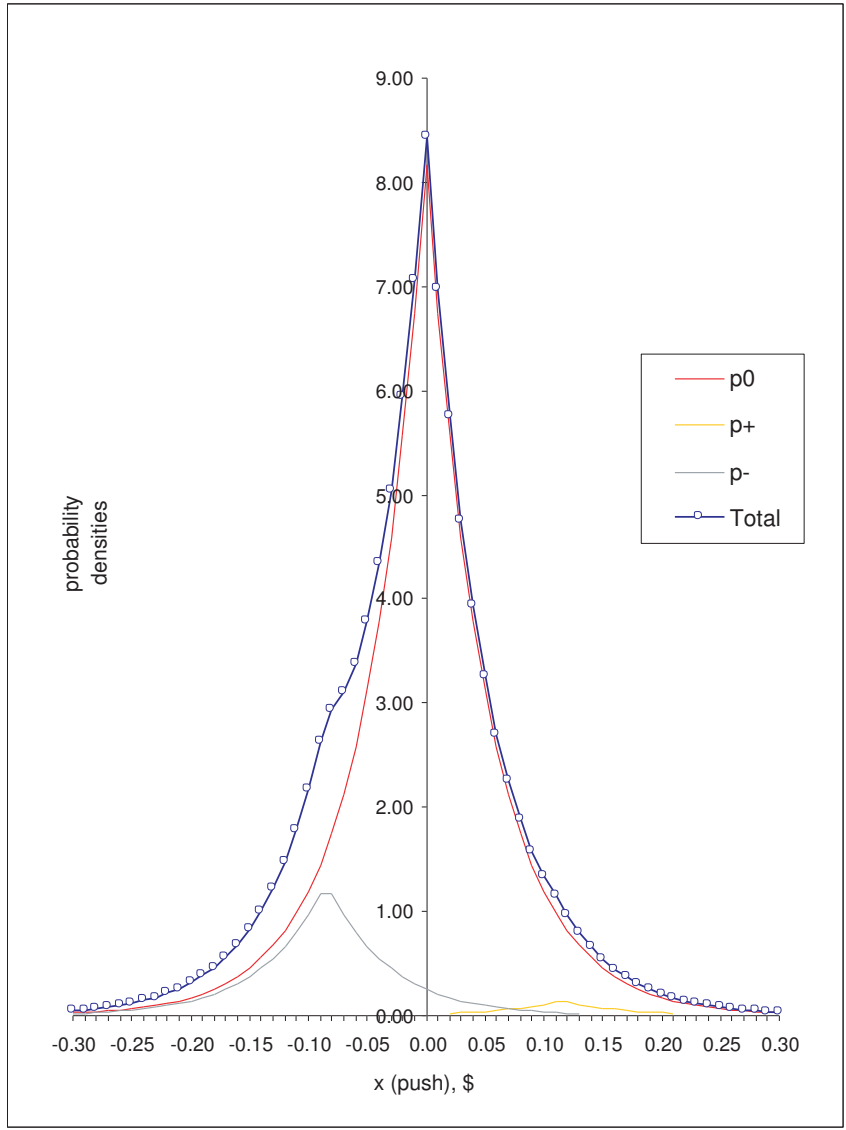}
 \end{center}
\caption{Components of conditional distribution ${\cal P} (y|\,x)$ at $x=0.1$.}
\end{figure}
\begin{figure}[h]
 \begin{center}
 \leavevmode
 \epsfysize=10cm
 \epsfbox{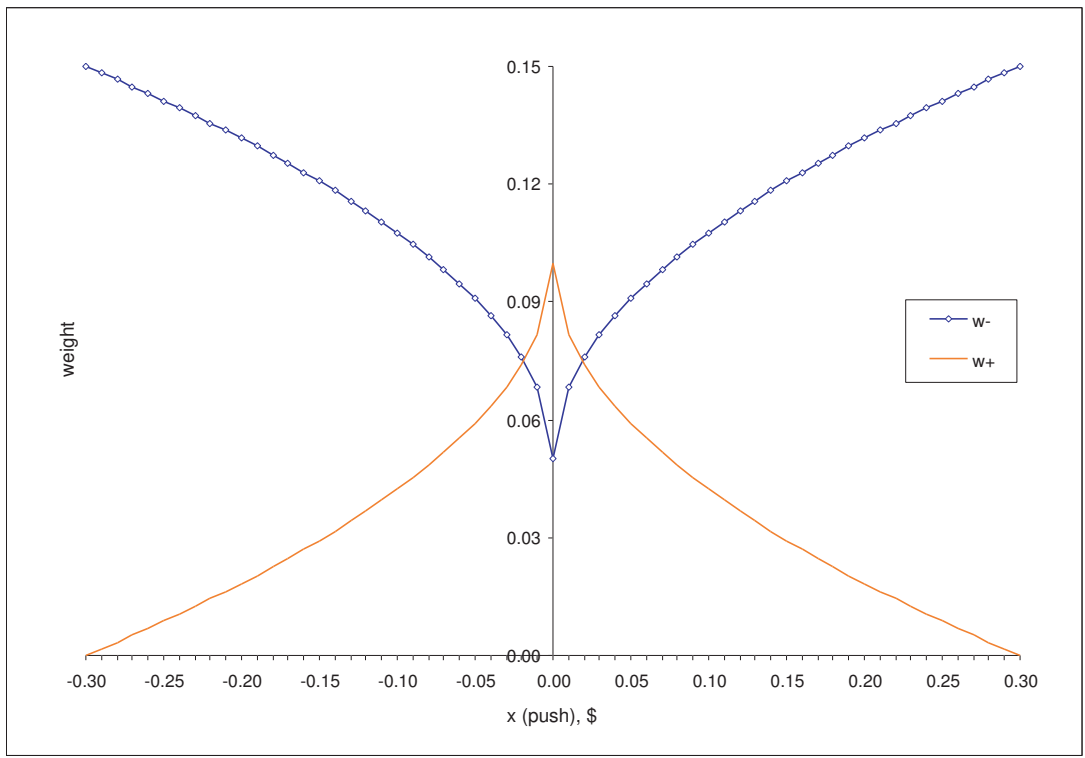}
 \end{center}
\caption{Weights $w^\pm$ of asymmetric components ${\cal P}^\pm (y|\,x)$.}
\end{figure}

\end{document}